\documentclass[12pt,final]{IEEEtran}
\usepackage{makeidx}  % allows for indexgeneration
\usepackage{amsfonts}
\usepackage{epsfig}
\usepackage{amsmath}
\usepackage{subfigure}
\usepackage[dutch,USenglish]{babel}
\numberwithin{equation}{section} \numberwithin{figure}{section}
\newtheorem{lemma}{Lemma}
\newtheorem{observation}{Observation}
\newtheorem{definition}{Definition}
\begin{document}
\title{On the Complexity of the Single Individual SNP Haplotyping Problem\thanks{Part of this research has been funded by the Dutch BSIK/BRICKS project.}}
\markboth{On the Complexity of the Single Individual SNP
Haplotyping Problem}{Cilibrasi \MakeLowercase{\textit{et al.}}}
\author{Rudi Cilibrasi\thanks{Rudi Cilibrasi is supported in part by NWO project 612.55.002, and by the IST Programme of the European
Community, under the PASCAL Network of Excellence,
IST-2002-506778. This publication only reflects the authors'
views.}, Leo van Iersel, Steven Kelk and John Tromp}
%
% \institute{Technische Universiteit Eindhoven (TU/e), Den Dolech 2, 5612 AX Eindhoven, Netherlands\\
% \email{l.j.j.v.iersel@tue.nl}\\
% \and
% Centrum voor Wiskunde en Informatica (CWI), Kruislaan 413, 1098 SJ Amsterdam, Netherlands \\
% \email{Rudi.Cilibrasi@cwi.nl, S.M.Kelk@cwi.nl, John.Tromp@cwi.nl}\\
% }
%
\maketitle              % typeset the title of the contribution
\begin{abstract}
We present several new results pertaining to haplotyping. These
results concern the combinatorial problem of reconstructing
haplotypes from incomplete and/or imperfectly sequenced haplotype
fragments. We consider the complexity of the problems
\emph{Minimum Error Correction} (MEC) and \emph{Longest Haplotype
Reconstruction} (LHR) for different restrictions on the input
data. Specifically, we look at the \emph{gapless} case, where
every row of the input corresponds to a gapless
haplotype-fragment, and the \emph{1-gap} case, where at most one
gap per fragment is allowed. We prove that MEC is APX-hard in the
1-gap case and still NP-hard in the gapless case. In addition, we
question earlier claims that MEC is NP-hard even when the input
matrix is restricted to being completely binary. Concerning LHR,
we show that this problem is NP-hard and APX-hard in the 1-gap
case (and thus also in the general case), but is polynomial time
solvable in the gapless case.
\end{abstract}
\section{Introduction}
If we abstractly consider the human genome as a string over the
nucleotide alphabet $\{ A, C, G, T \}$, it is widely known that
the genomes of any two humans have at more than 99\% of the sites
the same nucleotide. The sites at which variability is observed
across the human population are called \emph{Single Nucleotide
Polymorphisms} (SNPs), which are formally defined as the sites on
the human genome where, across the human population, two or more
nucleotides are observed and each such nucleotide occurs in at
least 5\% of the population. These sites, which occur (on average)
approximately once per thousand bases, capture the bulk of human
genetic variability; the string of nucleotides found at the SNP
sites of a human - the \emph{haplotype} of that individual - can
thus be thought of as a ``fingerprint'' for that individual.\\
\\
It has been observed that, for most SNP sites, only two
nucleotides are seen; sites where three or four nucleotides are
found are comparatively rare. Thus, from a combinatorial
perspective, a haplotype can be abstractly expressed as a string
over the alphabet $\{ 0,1 \}$. Indeed, the biologically-motivated
field of SNP and haplotype analysis has spawned a rich variety of
combinatorial problems, which are well described in surveys such
as \cite{bonizzoni} and \cite{halldorsson}.\\
\\
We focus on two such combinatorial problems, both variants of the
\emph{Single Individual Haplotyping Problem} (SIH), introduced in
\cite{lanciabafna}. SIH amounts to determining the haplotype of an
individual using (potentially) incomplete and/or imperfect
fragments of sequencing data. The situation is further complicated
by the fact that, being a \emph{diploid} organism, a human has two
versions of each chromosome; one each from the individual's mother
and father. Hence, for a given interval of the genome, a human has
two haplotypes. Thus, SIH can be more accurately described as
finding the two haplotypes of an individual given fragments of
sequencing data where the fragments potentially have read errors
and, crucially, where it is \emph{not} known which of the two
chromosomes each fragment was read from. We consider two
well-known variants of the problem: \emph{Minimum Error
Correction} (MEC), and \emph{Longest Haplotype Reconstruction}
(LHR).\\
\\
The input to these problems is a matrix $M$ of SNP fragments. Each
column of $M$ represents an SNP site and thus each entry of the
matrix denotes the (binary) choice of nucleotide seen at that SNP
location on that fragment. An entry of the matrix can thus either
be `0', `1' or a \emph{hole}, represented by `-', which denotes
lack of knowledge or uncertainty about the nucleotide at that
site. We use $M[i,j]$ to refer to the value found at row $i$,
column $j$ of $M$, and use $M[i]$ to refer to the $i$th row. Two
rows $r_1, r_2$ of the matrix \emph{conflict} if there exists a
column $j$ such that $M[r_1, j] \neq M[r_2, j]$ and $M[r_1,j],
M[r_2, j] \in \{0,1\}$.\\
\\
A matrix is \emph{feasible} iff the rows of the matrix can be
partitioned into two sets such that all rows
within each set are pairwise non-conflicting.\\
\\
The objective in MEC is to ``correct'' (or ``flip'') as few
entries of the input matrix as possible (i.e. convert 0 to 1 or
vice-versa) to arrive at a feasible matrix. The motivation behind
this is that all rows of the input matrix were sequenced from one
haplotype or the other, and that any deviation from
that haplotype occurred because of read-errors during sequencing.\\
\\
The problem LHR has the same input as MEC but a different
objective. Recall that the rows of a feasible matrix $M$ can be
partitioned into two sets such that all rows within each set are
pairwise non-conflicting. Having obtained such a partition, we can
reconstruct a haplotype from each set by merging all the rows in
that set together. (We define this formally later in Section
\ref{sec:lhr}.) With LHR the objective is to remove \emph{rows}
such that the resulting matrix is feasible and such that the sum
of the
lengths of the two resulting haplotypes is maximised.\\
\\
In the context of haplotyping, MEC and LHR have been discussed -
sometimes under different names - in papers such as
\cite{bonizzoni}, \cite{fasthare}, \cite{greenberg} and
(implicitly) \cite{lanciabafna}. One question arising from this
discussion is how the distribution of holes in the input data
affects computational complexity. To explain, let us first define
a \emph{gap} (in a string over the alphabet $\{0,1,-\}$) as a
maximal contiguous block of holes that is flanked on both sides by
non-hole values. For example, the string \texttt{---0010---} has
no gaps, \texttt{-0--10-111} has two gaps, and \texttt{-0-----1--}
has one gap. Two special cases of MEC and LHR that are considered
to be practically relevant are the ungapped case and the 1-gap
case. The ungapped variant is where every row of the input matrix
is ungapped, i.e. all holes appear at the start or end. In the
1-gap case every row has at most one gap.\\
In Section \ref{subsec:umec} we offer what we believe is the first
proof that Ungapped-MEC (and hence 1-gap MEC and also the general
MEC) is NP-hard. We do so by reduction from MAX-CUT. (As far as we
are aware, other claims of this result are based explicitly or
implicitly on results found in \cite{kleinberg}; as we discuss in
Section \ref{subsec:bmec}, we conclude that the results in
\cite{kleinberg} cannot be used for this purpose.)\\
\\
The NP-hardness of 1-gap MEC (and general MEC) follows immediately
from the proof that Ungapped-MEC is NP-hard. However, our
NP-hardness proof for Ungapped-MEC is not
approximation-preserving, and consequently tells us little about
the (in)approximability of Ungapped-MEC, 1-gap MEC and general
MEC. In light of this we provide (in Section \ref{subsec:gmec}) a
proof that 1-gap MEC is APX-hard, thus excluding (unless P=NP) the
existence of a \emph{Polynomial Time
Approximation Scheme} (PTAS) for 1-gap MEC (and general MEC.)\\
\\
We define (in Section \ref{subsec:bmec}) the problem
\emph{Binary-MEC}, where the input matrix contains no holes; as
far as we know the complexity of this problem is still -
intriguingly - open. We also consider a parameterised version of
binary-MEC, where the number of haplotypes is not fixed as two,
but is part of the input. We prove that this problem is NP-hard in
Section \ref{subsec:pbmec}. (In the Appendix we also prove an
``auxiliary'' lemma which, besides being interesting in its own
right, takes on a new significance in light of the open complexity
of
Binary-MEC.)\\
\\
In Section \ref{subsec:lhrpoly} we show that \emph{Ungapped-LHR}
is polynomial-time solvable and give a dynamic programming
algorithm for this which runs in time $O(n^{2}m+n^{3})$ for an $n
\times m$ input matrix. This improves upon the result of
\cite{lanciabafna} which also showed a polynomial-time algorithm
for Ungapped-LHR but
under the restricting assumption of non-nested input rows.\\
\\
We also prove, in Section \ref{subsec:lhrhard}, that LHR is
APX-hard (and thus also NP-hard) in the general case, by proving
the much stronger result that 1-gap LHR is APX-hard. This is the
first proof of hardness (for both 1-gap LHR and general LHR)
appearing in the literature. \footnote{In \cite{lanciabafna} there
is a claim, made very briefly, that LHR is NP-hard in general, but
it is not substantiated.}
\section{Minimum Error Correction (MEC)}
\label{sec:mec}
For a length-$m$ string $X \in \{0,1,-\}^m$, and a length-$m$
string $Y \in \{0,1\}^m$, we define $d(X,Y)$ as the number of
\emph{mismatches} between the strings i.e. positions where $X$ is
0 and $Y$ is 1, or vice-versa; holes do not contribute to the
mismatch count. Recall the definition of \emph{feasible} from
earlier; an alternative, and equivalent, definition (which we use
in the following proofs) is as follows. An $n \times m$ SNP matrix
$M$ is \emph{feasible} iff there exist two strings (haplotypes)
$H_1, H_2 \in \{0,1\}^m$,
such that for all rows r of M, $d( r, H_1) = 0$ or $d( r, H_2 )=0$.\\
\\
Finally, a \emph{flip} is where a 0 entry is converted to a 1, or
vice-versa. Flipping to or from holes is not allowed and the
haplotypes $H_1$ and $H_2$ may not contain holes.
\subsection{Ungapped-MEC}
\label{subsec:umec}
\noindent\textbf{Problem:} \emph{Ungapped-MEC}\\
\textbf{Input:} An ungapped SNP matrix $M$\\
\textbf{Output:} Ungapped-MEC(M), which we define as the smallest
number of flips needed to make $M$ feasible.\footnote{In
subsequent problem definitions we regard it as implicit that P(I)
represents the optimal output of a problem $P$ on input $I$.}\\
\begin{lemma}
\label{lem:mechard} Ungapped-MEC is NP-hard.\\
\end{lemma}
\begin{proof}
We give a polynomial-time reduction from MAX-CUT, which is the
problem of computing the size of a maximum cardinality cut in a
graph.\footnote{The reduction given here can easily be converted
into a Karp reduction from the decision version of MAX-CUT to the
decision version of Ungapped-MEC.} Let $G=(V,E)$ be the input to
MAX-CUT, where $E$ is undirected. (We identify, wlog, $V$ with
$\{1, 2,...,|V|\}$.) We construct an input matrix $M$ for
Ungapped-MEC with $2k|V| + |E|$ rows and $2|V|$ columns where $k =
2|E||V|$. We use $M_0$ to refer to the first $k|V|$ rows of $M$,
$M_1$ to refer to the second $k|V|$ rows of $M$, and $M_G$ to
refer to the remaining $|E|$ rows. $M_0$ consists of $|V|$
consecutive blocks of $k$ identical rows. Each row in the $i$-th
block (for $1 \leq i \leq |V|$) contains a $0$ at columns $2i-1$
and $2i$ and holes at all other columns. $M_1$ is defined similar
to $M_0$ with $1$-entries instead of $0$-entries. Each row of
$M_G$ encodes an edge from $E$: for edge $\{i,j\}$ (with $i<j$) we
specify that columns $2i-1$ and $2i$ contain 0s, columns $2j-1$
and $2j$ contain 1s, and for all $h \neq i, j$, column $2h-1$
contains 0 and column $2h$ contains 1. (See Figures
\ref{fig:mecgraph} and \ref{fig:mecmatrix} for an example of how
$M$ is constructed.)\\
\begin{figure}
\begin{centering}
\epsfig{file=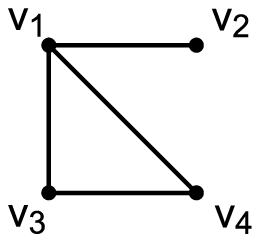} \caption{Example input to MAX-CUT (see
Lemma \ref{lem:mechard})} \label{fig:mecgraph}
\end{centering}
\end{figure}
\begin{figure}
\begin{centering}
\[
\begin{tabular}{rl}
$\left(
\begin{array}{cccccccc}
0 & 0 & - & - & - & - & - & - \\
- & - & 0 & 0 & - & - & - & - \\
- & - & - & - & 0 & 0 & - & - \\
- & - & - & - & - & - & 0 & 0 \\
1 & 1 & - & - & - & - & - & - \\
- & - & 1 & 1 & - & - & - & - \\
- & - & - & - & 1 & 1 & - & - \\
- & - & - & - & - & - & 1 & 1 \\
0 & 0 & 1 & 1 & 0 & 1 & 0 & 1 \\
0 & 0 & 0 & 1 & 1 & 1 & 0 & 1 \\
0 & 0 & 0 & 1 & 0 & 1 & 1 & 1 \\
0 & 1 & 0 & 1 & 0 & 0 & 1 & 1 %
\end{array}
\right) \hspace{-27pt}$ &
\begin{tabular}{l}
$\left.
\begin{array}{l}
\\
\\
\\
\\
\\
\\
\\
\\
\end{array}
\right\} 32$ copies \\
$\left.
\begin{array}{l}
\\
\\
\\
\\
\end{array}
\right\} M_G $\\
\end{tabular}
\end{tabular}
\]
\caption{Construction of matrix $M$ (from Lemma \ref{lem:mechard})
for graph in Figure \ref{fig:mecgraph}} \label{fig:mecmatrix}
\end{centering}
\vspace{-12pt}
\end{figure}
\\
Suppose $t$ is the largest cut possible in $G$ and $s$ is the
minimum number of flips needed to make $M$ feasible. We claim that
the following holds:
\begin{equation}
\label{maxcut} s=|E|(|V|-2)+2(|E|-t).
\end{equation}
From this $t$, the optimal solution of MAX-CUT, can easily be
computed. First, note that the solution to Ungapped-MEC(M) is
trivially upperbounded by $|V||E|$. This follows because we could
simply flip every 1 entry in $M_G$ to 0; the resulting overall
matrix would be feasible because we could just take $H_1$ as the
all-0 string and $H_2$ as the all-1 string. Now, we say a
haplotype $H$ has the \emph{double-entry} property if, for all
odd-indexed positions (i.e. columns) $j$ in $H$, the entry at
position $j$ of $H$ is the same as the entry at position $j+1$. We
argue that a minimal number of feasibility-inducing flips will
\emph{always} lead to two haplotypes $H_1, H_2$ such that both
haplotypes have the double-entry property and, further, $H_1$ is
the bitwise complement of $H_2$. (We describe such a pair of
haplotypes as \emph{partition-encoding}.) This is because, if
$H_1, H_2$ are not partition-encoding, then at least $k > |V||E|$
(in contrast with zero) entries in $M_0$ and/or $M_1$ will have to
be flipped, meaning this strategy is doomed to begin with.\\
\\
Now, for a given partition-encoding pair of haplotypes, it follows
that - for each row in $M_G$ - we will have to flip either $|V|-2$
or $|V|$ entries to reach its nearest haplotype. This is because,
irrespective of which haplotype we move a row to, the $|V|-2$
pairs of columns \emph{not} encoding end-points (for a given row)
will always cost 1 flip each to fix. Then either 2 or 0 of the 4
``endpoint-encoding'' entries will also need to be flipped; 4
flips will never be necessary because then the row could move to
the other haplotype, requiring no extra flips. Ungapped-MEC thus
maximises the number of rows which require $|V|-2$ rather than
$|V|$ flips. If we think of $H_1$ and $H_2$ as encoding a
partition of the vertices of $V$ (i.e. a vertex $i$ is on one side
of the partition if $H_1$ has 1s in columns $2i-1$ and $2i$, and
on the other side if $H_2$ has 1s in those columns), it follows
that each row requiring $|V|-2$ flips corresponds to a cut-edge in
the vertex partition defined by $H_1$ and $H_2$. The expression
(\ref{maxcut}) follows.\\
\end{proof}
\subsection{1-gap MEC}
\label{subsec:gmec}
\noindent\textbf{Problem:} \emph{1-gap MEC}\\
\textbf{Input:} SNP matrix $M$ with at most 1 gap per row\\
\textbf{Output:} The smallest number of flips needed to make $M$ feasible.\\
\\
To prove that 1-gap MEC is APX-hard (and therefore also NP-hard)
we will give an L-reduction\footnote{An L-reduction is a specific
type of \emph{approximation-preserving} reduction, first
introduced in \cite{lreduc}. If there exists an L-reduction from a
problem X to a problem Y, then a PTAS for Y can be used to build a
PTAS for X. Conversely, if there exists an L-reduction from X to
Y, and X is APX-hard, so is Y. See (for example) \cite{sched} for
a succinct discussion of this.} from CUBIC-MIN-UNCUT, which is the
problem of finding the minimum number of edges that have to be
removed from a 3-regular graph in order to make it bipartite. Our
first goal is thus to prove the APX-hardness of CUBIC-MIN-UNCUT,
which itself will be proven using an L-reduction from the APX-hard problem CUBIC-MAX-CUT.\\
\\
To aid the reader, we reproduce here the definition of an
L-reduction.\\
\begin{definition}
(Papadimitriou and Yannakakis \cite{lreduc}) Let A and B be two
optimisation problems. An \emph{L-reduction} from A to B is a pair
of functions R and S, both computable in polynomial time, such
that for any instance I of A with optimum cost Opt(I), R(I) is an
instance of B with optimum cost Opt(R(I)) and for every
feasible\footnote{Note that \emph{feasible} in this context has a
different meaning to \emph{feasible} in the context of SNP
matrices.} solution s of R(I), S(s) is a feasible solution of I
such that:
\begin{equation}
\label{eq:L1}
Opt(R(I)) \leq \alpha Opt(I),
\end{equation}
for some positive constant $\alpha$ and:
\begin{equation}
\label{eq:L2}
|Opt(I) - c(S(s))| \leq \beta |Opt(R(I))-c(s)|,
\end{equation}
for some positive constant $\beta$, where c(S(s)) and c(s)
represent the costs of S(s) and s, respectively.\\
\end{definition}
\begin{observation}
\label{obs:mincuthard} CUBIC-MIN-UNCUT is APX-hard.\\
\end{observation}
\begin{proof}
We give an L-reduction from CUBIC-MAX-CUT, the problem of finding
the maximum cardinality of a cut in a 3-regular graph. (This
problem is shown to be APX-hard in \cite{alimontikann}; see also
\cite{bermankarpinski}.) Let $G=(V,E)$
be the input to CUBIC-MAX-CUT.\\
\\
Note that CUBIC-MIN-UNCUT is the ``complement'' of CUBIC-MAX-CUT,
as expressed by the following relationship:
\begin{equation}
\begin{array}{l}
\label{eq:duality} \text{\emph{CUBIC-MAX-CUT(G)}}\\
= |E| - \text{\emph{CUBIC-MIN-UNCUT(G)}}.
\end{array}
\end{equation}
To see why this holds, note that for every cut $C$, the removal of
the edges $E \setminus C$ will lead to a bipartite graph. On the
other hand, given a set of edges $E'$ whose removal makes $G$
bipartite, the complement is not necessarily a cut. However, given
a bipartition induced by the removal of $E'$, the edges from the
original graph that cross this bipartition form a cut $C'$, such
that $|C'| \geq |E \setminus E'|$. This proves (\ref{eq:duality}),
and the mapping (just described) from $E'$ to $C'$ is the mapping
we use in the L-reduction.\\
\\
Now, note that property (\ref{eq:L1}) of the L-reduction is easily
satisfied (taking $\alpha=1$) because the optimal value of
CUBIC-MIN-UNCUT is always less than or equal to the optimal value
of CUBIC-MAX-CUT. This follows from the combination of
(\ref{eq:duality}) with the fact that a maximum cut in a 3-regular
graph always contains at least $2/3$ of the edges: if a vertex has
less than two incident edges in the cut then we can get a larger
cut by moving this vertex to the other side of the partition.\\
\\
To see that property (\ref{eq:L2}) of the L-reduction is easily
satisfied (taking $\beta = 1$), let $E'$ be any set of edges whose
removal makes $G$ bipartite. Property (\ref{eq:L2}) is satisfied
because $E'$ gets mapped to a cut $C'$, as defined above, and
combined with (\ref{eq:duality}) this gives:
\begin{equation}
\begin{array}{l}
\text{\emph{CUBIC-MAX-CUT(G)}} - |C'|\\
\leq \text{\emph{CUBIC-MAX-CUT(G)}} - |E \setminus E'| \\
= |E'| - \text{\emph{CUBIC-MIN-UNCUT(G)}}.
\end{array}
\end{equation}
This completes the L-reduction from CUBIC-MAX-CUT to
CUBIC-MIN-UNCUT, proving the APX-hardness of CUBIC-MIN-UNCUT.\\
\end{proof}
We also need the following observation.\\
\begin{observation}
\label{obs:orient} Let $G = (V,E)$ be an undirected, 3-regular
graph. Then we can find, in polynomial time, an orientation of the
edges of $G$ so that each vertex has either in-degree 2 and
out-degree 1 (``in-in-out'') or
out-degree 2 and in-degree 1 (``out-out-in'').\\
\end{observation}
\begin{proof}
(We assume that $G$ is connected; if $G$ is not connected, we can
apply the following argument to each component of $G$ in turn, and
the overall result still holds.) Every cubic graph has an even
number of vertices, because every graph must have an even number
of odd-degree vertices. We add an arbitrary perfect matching to
the graph, which may create multiple edges. The graph is now
4-regular and therefore has an Euler tour. We direct the edges
following the Euler-tour; every vertex is now in-in-out-out. If we
remove the perfect matching edges we added, we are left with an
oriented version of $G$ where every vertex is in-in-out or
out-out-in. This can all be done in polynomial time.\\
\end{proof}
\begin{lemma}
\label{apxhard} 1-gap MEC is APX-hard\\
\end{lemma}
\begin{proof}
We give a reduction from CUBIC-MIN-UNCUT. Consider an arbitrary
3-regular graph $G = (V,E)$ and orient the edges as described in
Observation \ref{obs:orient} to obtain an oriented version of $G$,
$\overrightarrow{G} = (V, \overrightarrow{E})$, where each vertex
is either in-in-out or out-out-in. We construct an $|E| \times
|V|$ input matrix $M$ for 1-gap MEC as follows. The columns of $M$
correspond to the vertices of $\overrightarrow{G}$ and every row
of $M$ encodes an oriented edge of $\overrightarrow{G}$; it has a
$0$ in the column corresponding to the tail of the edge (i.e. the
vertex from which the edge leaves), a $1$ in the column
corresponding to the head of the edge,
and the rest holes.\\
\\
We prove the following:
\begin{equation}
\label{eq:uncutmec} \text{\emph{CUBIC-MIN-UNCUT(G)}} =
\text{\emph{1-gap MEC(M)}}.
\end{equation}
We first prove that:
\begin{equation}
\label{eq:uncutmec2} \text{\emph{1-gap MEC(M)}} \leq
\text{\emph{CUBIC-MIN-UNCUT(G)}}.
\end{equation}
To see this, let $E'$ be a minimal set of edges whose removal
makes $G$ bipartite, and let $|E'| = k$. Let $B = (L \cup R, E
\setminus E')$ be the bipartite graph (with bipartition $L \cup
R$) obtained from $G$ by removing the edges $E'$. Let $H_1$
(respectively, $H_2$) be the haplotype that has 1s in the columns
representing vertices of $L$ (respectively, $R$) and 0s elsewhere.
It is possible to make $M$ feasible with $k$ flips, by the
following process: for each edge in $E'$, flip the 0 bit in the
corresponding row of $M$ to 1. For each row r of M it is now
true that $d(r, H_1) = 0$ or $d(r,H_2) = 0$, proving the feasibility of $M$.\\
\\
The proof that,
\begin{equation}
\label{eq:uncutmec3} \text{\emph{CUBIC-MIN-UNCUT(G)}}\leq
\text{\emph{1-gap MEC(M)}},
\end{equation}
is more subtle. Suppose we can render $M$ feasible using $j$
flips, and let $H_1$ and $H_2$ be any two haplotypes such that,
after the $j$ flips, each row of $M$ is distance 0 from either
$H_1$ or $H_2$. If $H_1$ and $H_2$ are bitwise complementary then
we can make $G$ bipartite by removing an edge whenever we had to
flip a bit in the corresponding row. The idea is, namely, that the
1s in $H_1$ (respectively, $H_2$) represent the vertices $L$
(respectively, $R$) in the resulting bipartition $L \cup R$.\\
\\
However, suppose the two haplotypes $H_1$ and $H_2$ are not
bitwise complementary. In this case it is sufficient to
demonstrate that there also exists bitwise complementary
haplotypes $H'_1$ and $H'_2$ such that, after $j$ (or fewer)
flips, every row of $M$ is distance 0 from either $H'_1$ or
$H'_2$. Consider thus a column of $H_1$ and $H_2$ where the two
haplotypes are not complementary. Crucially, the orientation of
$\overrightarrow{G}$ ensures that every column of $M$ contains
\emph{either} one $1$ and two $0$s \emph{or} two $1$s and one $0$
(and the rest holes). A simple case analysis shows that, because
of this, we can always change the value of one of the haplotypes
in that column, without increasing the number of flips. (The
number of flips might decrease.) Repeating this process for all
columns of $H_1$ and $H_2$ where the same value is observed thus
creates complementary haplotypes $H'_1$ and $H'_2$, and - as
described in the previous paragraph - these haplotypes then
determine which edges of $G$ should be removed to make $G$
bipartite. This completes the proof of (\ref{eq:uncutmec}).\\
\\
The above reduction can be computed in polynomial time and is an
L-reduction. From (\ref{eq:uncutmec}) it follows directly that
property (\ref{eq:L1}) of an L-reduction is satisfied with
$\alpha=1$. Property (\ref{eq:L2}), with $\beta=1$, follows from
the proof of (\ref{eq:uncutmec3}), combined with
(\ref{eq:uncutmec}). Namely, whenever we use (say) $t$ flips to
make $M$ feasible, we can find $s \leq t$ edges of $G$ that can be
removed to make $G$ bipartite. Combined with (\ref{eq:uncutmec})
this gives:
\begin{equation}
\begin{array}{l}
|\text{\emph{CUBIC-MIN-UNCUT(G)}} - s |\\
\leq | \text{\emph{1-gap MEC(M)}} - t |.
\end{array}
\end{equation}
\end{proof}
\subsection{Binary-MEC}
\label{subsec:bmec}
From a mathematical point of view it is interesting to determine
whether MEC stays NP-hard when the input matrix is
further restricted. Let us therefore define the following problem.\\
\\
\textbf{Problem:} \emph{Binary-MEC}\\
\textbf{Input:} An SNP matrix $M$ that does not contain any holes\\
\textbf{Output:} As for Ungapped-MEC\\
\\
Like all optimisation problems, the problem Binary-MEC has
different variants, depending on how the problem is defined. The
above definition is technically speaking the \emph{evaluation}
variant of the Binary-MEC problem\footnote{ See \cite{ausiello}
for a more detailed explanation of terminology in this area.}.
Consider the closely-related \emph{constructive} version:\\
\\
\textbf{Problem:} \emph{Binary-Constructive-MEC}\\
\textbf{Input:} An SNP matrix $M$ that does not contain any holes\\
\textbf{Output:} For an input matrix $M$ of size $n \times m$, two
haplotypes $H_1, H_2 \in \{0,1\}^m$ minimizing:
\begin{equation}
\label{eq:witsum} D_M(H_1, H_2) = \sum_{\text{rows r of M}} \min(
d(r,H_1), d(r, H_2) ).
\end{equation}
In the appendix, we prove that Binary-Constructive-MEC is
polynomial-time Turing interreducible with its evaluation
counterpart, Binary-MEC. This proves that Binary-Constructive-MEC
is solvable in polynomial-time iff Binary-MEC is solvable in
polynomial-time. We mention this correspondence because, when
expressed as a constructive problem, it can be seen that MEC is in
fact a specific type of \emph{clustering} problem, a topic of
intensive study in the literature. More specifically, we are
trying to find two representative ``median'' (or ``consensus'')
strings such that the sum, over all input strings, of the distance
between each input string and its nearest median, is minimised.
This interreducibility is potentially useful because we now argue,
in contrast to claims in the existing literature, that the
complexity
of Binary-MEC / Binary-Constructive-MEC is actually still open.\\
\\
To elaborate, it is claimed in several papers (e.g. \cite{alon})
that a problem equivalent to Binary-Constructive-MEC is NP-hard.
Such claims inevitably refer to the seminal paper
\emph{Segmentation Problems} by Kleinberg, Papadimitriou, and
Raghavan (KPR), which has appeared in multiple different forms
since 1998 (e.g. \cite{kleinberg}, \cite{kleinbergEco} and
\cite{kleinberg2004}.) However, the KPR papers actually discuss
two superficially similar, but essentially different, problems:
one problem is essentially equivalent to Binary-Constructive-MEC,
and the other is a more general (and thus, potentially, a more
difficult) problem.\footnote{In this more general problem, rows
and haplotypes are viewed as vectors and the distance between a
row and a haplotype is their dot product. Further, unlike
Binary-Constructive-MEC, this problem allows entries of the input
matrix to be drawn arbitrarily from $\mathbb{R}$. This extra
degree of freedom - particularly the ability to simultaneously use
positive, negative and zero values in the input matrix - is what
(when coupled with a dot product distance measure) provides the
ability to encode NP-hard problems.} Communication with the
authors \cite{christos} has confirmed that they have no proof of
hardness for the former problem i.e. the problem that is
essentially equivalent to Binary-Constructive-MEC.\\
\\
Thus we conclude that the complexity of Binary-Constructive-MEC /
Binary-MEC is still open. From an approximation viewpoint the
problem has been quite well-studied; the problem has a
\emph{Polynomial Time Approximation Scheme} (PTAS) because it is a
special form of the \emph{Hamming 2-Median Clustering Problem},
for which a PTAS is demonstrated in \cite{li}. Other approximation
results appear in \cite{kleinberg}, \cite{alon},
\cite{kleinberg2004}, \cite{geometric} and a heuristic for a
similar (but not identical) problem appears in \cite{fasthare}. We
also know that, if the number of haplotypes to be found is
specified as part of the input (and not fixed as 2), the problem
becomes NP-hard; we prove this in the following section. Finally,
it may also be relevant that the ``geometric'' version of the
problem (where rows of the input matrix are not drawn from
$\{0,1\}^m$ but from $\mathbb{R}^{m}$, and Euclidean distance is
used instead of Hamming distance) is also open from a complexity
viewpoint \cite{geometric}. (However, the version using
Euclidean-distance-squared \emph{is} known to be NP-hard
\cite{drineas}.)
\subsection{Parameterised Binary-MEC}
\label{subsec:pbmec}
Let us now consider a generalisation of the problem Binary-MEC,
where the number of haplotypes is not fixed as two, but part of
the input.\\
\\
\textbf{Problem:} \emph{Parameterised-Binary-MEC (PBMEC)}\\
\textbf{Input:} An SNP matrix $M$ that contains no holes, and $k \in \mathbb{N} \setminus \{0\}$\\
\textbf{Output:} The smallest number of flips needed to make $M$
feasible under $k$ haplotypes.\\
The notion of \emph{feasible} generalises easily to $k \geq 1$
haplotypes: an SNP matrix $M$ is \emph{feasible} under $k$
haplotypes if $M$ can be partitioned into $k$ segments such that
all the rows within each segment are pairwise non-conflicting. The
definition of $D_{M}$ also generalises easily to $k$ haplotypes;
we define $D_{M, k}(H_1, H_2, ..., H_k)$ as:
\begin{equation}
\label{eq:kmedsum} \sum_{\text{rows r of M}} \min(d(r,H_1), d(r,
H_2), ..., d(r,H_k) ).
\end{equation}
We define $OptTuples(M,k)$ as the set of unordered optimal
$k$-tuples of haplotypes for $M$ i.e. those $k$-tuples of
haplotypes which have a $D_{M,k}$ score equal to PBMEC$(M,k)$.\\
\begin{lemma}
\label{lem:pbmechard}
PBMEC is NP-hard\\
\end{lemma}
\begin{proof}
We reduce from the NP-hard problem MINIMUM-VERTEX-COVER. Let
$G=(V,E)$ be an undirected graph. A subset $V' \subseteq V$ is
said to \emph{cover} an edge $(u,v) \in E$ iff $u \in V'$ or $v
\in V'$. A \emph{vertex cover} of an undirected graph $G = (V,E)$
is a subset $U$ of the vertices such that every edge in $E$ is
covered by $U$. MINIMUM-VERTEX-COVER is the problem of, given a
graph $G$, computing the size of
a minimum cardinality vertex cover $U$ of $G$.\\
\\
Let $G = (V,E)$ be the input to MINIMUM-VERTEX-COVER. We construct
an SNP matrix $M$ as follows. $M$ has $|V|$ columns and
$3|E||V|+|E|$ rows. We name the first $3|E||V|$ rows $M_0$ and the
remaining $|E|$ rows $M_{G}$. $M_0$ is the matrix obtained by
taking the $|V| \times |V|$ identity matrix (i.e. 1s on the
diagonal, 0s everywhere else) and making $3|E|$ copies of each
row. Each row in $M_G$ encodes an edge of $G$: the row has
1-entries at the endpoints of the edge, and the rest of the row is
0. We argue shortly that, to compute the size of the smallest
vertex cover in $G$, we call PBMEC($M,k$) for increasing values of
$k$ (starting with $k=2$) until we first encounter a $k$ such
that:
\begin{equation}
\label{eq:kmed} PBMEC(M,k) = 3|E|(|V|-(k-1)) + |E|.
\end{equation}
Once the smallest such $k$ has been found, we can output that the
size of the smallest vertex cover in $G$ is $k-1$. (Actually, if
we haven't yet found a value $k < |V|-2$ satisfying the above
equation, we can check by brute force in polynomial-time whether
$G$ has a vertex cover of size $|V|-3$, $|V|-2$, $|V|-1$, or
$|V|$. The reason for wanting to ensure that PBMEC($M,k$) is not
called with $k \geq |V|-2$ is explained later in the
analysis.\footnote{Note that, should we wish to build a Karp
reduction from the decision version of MINIMUM-VERTEX-COVER to the
decision version of PBMEC, it is not a problem to make this brute
force checking fit into the framework of a Karp reduction. The
Karp reduction can do the brute force checking itself and use
trivial inputs to the decision
version of PBMEC to communicate its ``yes'' or ``no'' answer.})\\
\\
It remains only to prove that (for $k < |V|-2$)
(\ref{eq:kmed}) holds iff $G$ has a vertex cover of size $k-1$.\\
\\
To prove this we need to first analyze $OptTuples(M_{0},k)$.
Recall that $M_0$ was obtained by duplicating the rows of the $|V|
\times |V|$ identity matrix. Let $I_{|V|}$ be shorthand for the
$|V| \times |V|$ identity matrix. Given that $M_0$ is simply a
``scaled up'' version of $I_{|V|}$, it follows that:
\begin{equation}
OptTuples(M_0,k) = OptTuples(I_{|V|},k).
\end{equation}
Now, we argue that all the $k$-tuples in $OptTuples(I_{|V|},k)$
(for $k < |V|-2$) have the following form: one haplotype from the
tuple contains only 0s, and the remaining $k-1$ haplotypes from
the tuple each have precisely one entry set to 1. Let us name such
a $k$-tuple
a \emph{candidate} tuple.\\
\begin{figure}
\begin{centering}
\epsfig{file=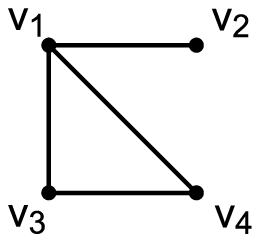}
\caption{Example input graph to
MINIMUM-VERTEX-COVER (see Lemma \ref{lem:pbmechard})}
\label{fig:pbmecgraph}
\end{centering}
\end{figure}
\begin{figure}
\begin{centering}
\[
\begin{tabular}{ll}
$\left(
\begin{array}{cccc}
1 & 0 & 0 & 0 \\
0 & 1 & 0 & 0 \\
0 & 0 & 1 & 0 \\
0 & 0 & 0 & 1 \\
1 & 1 & 0 & 0 \\
1 & 0 & 1 & 0 \\
1 & 0 & 0 & 1 \\
0 & 0 & 1 & 1 \\
\end{array}
\right)$ \hspace{-28pt} &
\begin{tabular}{l}
$\left.
\begin{array}{c}
\\
\\
\\
\\
\end{array}
\right\} 12$ copies \\
$\left.
\begin{array}{c}
\\
\\
\\
\\
\end{array}
\right\} M_G $\\
\end{tabular}
\end{tabular}
\]
\end{centering}
\caption{Construction of matrix $M$ for graph from Figure
\ref{fig:pbmecgraph}}
\end{figure}
\\
First, note that $PBMEC(I_{|V|},k) \leq |V|-(k-1)$, because
$|V|-(k-1)$ is the value of the $D$ measure - defined in
(\ref{eq:kmedsum}) - under any candidate tuple. Secondly, under an
arbitrary $k$-tuple there can be at most $k$ rows of $I_{|V|}$
which contribute 0 to the $D$ measure. However, if precisely $k$
rows of $I_{|V|}$ contribute 0 to the $D$ measure (i.e. every
haplotype has precisely one entry set to 1, and the haplotypes are
all distinct) then there are $|V|-k$ rows which each contribute 2
to the $D$ measure; such a $k$-tuple cannot be optimal because it
has a $D$ measure of $2(|V|-k) > |V|-(k-1)$. So we reason that at
most $k-1$ rows contribute 0 to the $D$ measure. In fact,
\emph{precisely} $k-1$ rows must contribute 0 to the $D$ measure
because, otherwise, there would be at least $|V|-(k-2)$ rows
contributing at least 1, and this is not possible because
$PBMEC(I_{|V|},k) \leq |V|-(k-1)$. So $k-1$ of the haplotypes
correspond to rows of $I_{|V|}$, and the remaining $|V|-(k-1)$
rows of $I_{|V|}$ must each contribute 1 to the $D$ measure. But
the only way to do this (given that $|V|-(k-1) > 2$) is to make
the $k$th haplotype the haplotype where every entry is 0. Hence:
\begin{equation}
PBMEC(I_{|V|},k) = |V|-(k-1)
\end{equation}
and:
\begin{equation}
PBMEC(M_0,k) = 3|E|(|V|-(k-1)).
\end{equation}
$OptTuples(I_{|V|},k)$ ($= OptTuples(M_0,k)$) is, by extension,
precisely the set of candidate $k$-tuples.\\
\\
The next step is to observe that $OptTuples(M,k) \subseteq
OptTuples(M_0,k)$. To see this, suppose (by way of contradiction)
that it is not true, and there exists a $k$-tuple $H^{*} \in
OptTuples(M,k)$ that is not in $OptTuples(M_0,k)$. But then
replacing $H^{*}$ by any $k$-tuple out of $OptTuples(M_0,k)$ would
reduce the number of flips needed in $M_0$ by at least $3|E|$, in
contrast to an increase in the number of flips needed in $M_{G}$
of at most $2|E|$, thus leading to an overall reduction in the
number of flips; contradiction! (The $2|E|$ figure is the number
of flips
required to make all rows in $M_G$ equal to the all-0 haplotype.)\\
\\
Because $OptTuples(M,k) \subseteq OptTuples(M_0,k)$, we can
restrict our attention to the $k$-tuples in $OptTuples(M_0,k)$.
Observe that there is a natural 1-1 correspondence between the
elements of $OptTuples(M_0,k)$ and all size $k-1$ subsets of $V$:
a vertex $v \in V$ is in the subset corresponding to $H^{*} \in
OptTuples(M_0,k)$ iff one of the haplotypes in $H^{*}$ has a 1 in
the column corresponding to vertex $v$.\\
\\
Now, for a $k$-tuple $H^{*} \in OptTuples(M_0,k)$ we let $Cov( G,
H^{*} )$ be the set of edges in $G$ which are covered by the
subset of $V$ corresponding to $H^{*}$. (Thus, $|Cov(G,
H^{*})|=|E|$ iff $H^{*}$ represents a vertex cover of $G$.) It is
easy to check that, for $H^{*} \in OptTuples(M_0,k)$:
\begin{equation}
\begin{array}{lll}
D_{M,k}( H^{*} )&{}={}&3|E|(|V|-(k-1))\\
&&{+}\:|Cov(G, H^{*})|\\
&&{+}\:2( |E| - |Cov(G,H^{*}| ) \\
&{}={}&3|E|(|V|-(k-1))\\
&&{+}\:2|E| - |Cov(G, H^{*})|.\nonumber
\end{array}
\end{equation}
Hence, for $H^{*} \in OptTuples(M_0,k)$, $D_{M,k}(H^{*})$ equals
$3|E|(|V|-(k-1)) + |E|$ iff $H^{*}$ represents a size $k-1$ vertex
cover of $G$.\\
\end{proof}
\section{Longest Haplotype Reconstruction (LHR)}
\label{sec:lhr} \setcounter{equation}{0} Suppose an SNP matrix $M$
is feasible. Then we can partition the rows of $M$ into two sets,
$M_l$ and $M_r$, such that the rows within each set are pairwise
non-conflicting. (The partition might not be unique.) From $M_i$
($i \in \{l,r\}$) we can then build a haplotype $H_i$ by combining
the rows of $M_i$ as follows: The $j$th column of $H_i$ is set to
1 if at least one row from $M_i$ has a 1 in column $j$, is set to
0 if at least one row from $M_i$ has a 0 in column $j$, and is set
to a hole if all rows in $M_i$ have a hole in column $j$. Note
that, in contrast to MEC, this leads to haplotypes that
potentially contain holes. For example, suppose one side of the
partition contains rows \texttt{10--, -0--} and \texttt{---1};
then the haplotype we get from this is \texttt{10-1}. We define
the \emph{length} of a haplotype $H$, denoted as $|H|$, as the
number of positions where it does not contain a hole; the
haplotype \texttt{10-1} thus has length 3, for example. Now, the
objective with LHR is to remove \emph{rows} from $M$ to make it
feasible but also such that the sum of the lengths of the two
resulting haplotypes is maximised. We define the function LHR(M)
(which gives a natural number as output) as the largest value this
sum-of-lengths value can take,
ranging over all feasibility-inducing row-removals and subsequent partitions.\\
\\
In Section \ref{subsec:lhrpoly} we provide a polynomial-time
dynamic programming algorithm for the ungapped variant of LHR,
Ungapped-LHR. In Section \ref{subsec:lhrhard} we show that LHR
becomes APX-hard and NP-hard when at most one gap per input row is
allowed, automatically also proving the hardness of LHR in the
general case.
\subsection{A polynomial-time algorithm for Ungapped-LHR}
\label{subsec:lhrpoly}
\noindent\textbf{Problem:} \emph{Ungapped-LHR}\\
\textbf{Input: } An ungapped SNP matrix $M$\\
\textbf{Output: } The value LHR(M), as defined above\\
\\
The LHR problem for ungapped matrices was proved to be
polynomial-time solvable by Lancia et. al in \cite{lanciabafna},
but only with the genuine restriction that no fragments are
included in other fragments. Our algorithm improves this in the
sense that it works for all ungapped input matrices; our algorithm
is similar in style to the algorithm that solves
MFR\footnote{Minimum Fragment Removal: in this problem the
objective is not to maximise the length of the haplotypes, but to
minimise the number of rows removed} in the ungapped case by Bafna
et. al. in \cite{bafna2005}. Note that our dynamic-programming
algorithm computes Ungapped-LHR(M) but it can easily be adapted to
generate the rows that must be removed (and subsequently, the
partition that must be made) to achieve this value.\\
\begin{lemma}
Ungapped-LHR can be solved in time $O(n^{2}m + n^{3})$\\
\end{lemma}
\begin{proof}
Let $M$ be the input to Ungapped-LHR, and assume the matrix has
size $n \times m$. For row $i$ define $l(i)$ as the leftmost
column that is not a hole and define $r(i)$ as the rightmost
column that is not a hole. The rows of $M$ are ordered such that
$l(i)\leq l(j)$ if $i<j$. Define the matrix $M_{i}$ as the matrix
consisting of the first $i$ rows of $M$ and two extra rows at the
top: row $0$ and row $-1$, both consisting of all holes. Define
$W(i)$ as the set of rows $j<i$ that are not in conflict with row
$i$.\\
\\
For $h,k\leq i$ and $h,k\geq -1$ and $r(h)\leq r(k)$ define
$D[h,k;i]$ as the maximum sum of lengths of two haplotypes such
that:
\begin{itemize}
\item each haplotype is built up as a combination of rows from
$M_i$ (in the sense explained above);
\item each row from $M_{i}$
can be used to build at most one haplotype (i.e. it cannot be used
for both haplotypes);
\item row $k$ is one of the rows used to build a haplotype and among such rows maximises $%
r(\cdot )$; \item row $h$ is one of the rows used to build the
haplotype for which $k$ is not used and among such rows maximises
$r(\cdot )$.\\
\end{itemize}
The optimal solution of the problem, $LHR(M)$, is given by:
\begin{equation}
\max_{h,k|r(h)\leq r(k)}D[h,k;n].
\end{equation}
This optimal solution can be calculated by starting with
$D[h,k,0]=0$ for $h,k\in {-1,0}$ and using the following recursive
formulas. We distinguish three different cases, the first is that
$h,k<i$. Under these circumstances:
\begin{equation}
\label{eq:lhr1} D[h,k;i]=D[h,k;i-1].
\end{equation}
This is because:
\begin{itemize}
\item if $r(i)>r(k)$: row $i$ cannot be used for the haplotype
that row $k$ is used for, because row $k$ has maximal $r(\cdot )$
among all rows that are used for a haplotype; \item if $r(i)\leq
r(k)$: row $i$ cannot increase the length of the haplotype that
row $k$ is used for (because also $l(i)\geq l(k)$);
\item the same arguments hold for $h$.\\
\end{itemize}
The second case is when $h=i$; $D[i,k;i]$ is equal to:
\begin{equation}
\label{eq:lhr}
\max_{\substack{ j\in W(i),\text{ \ ~}j\neq k \\ r(j)\leq r(i)}}%
D[j,k;i-1]+f(i,j).
\end{equation}
Where $f(i,j)=r(i)-\max \{r(j),l(i)-1\}$ is the increase of the
haplotype's length. Equation (\ref{eq:lhr}) results from the following.
The definition of $D[i,k;i]$ says that row $%
i$ has to be used for the haplotype for which $k$ is not used and
amongst such rows maximises $r(\cdot )$. Therefore, the optimal
solution is achieved by adding row $i$ to some solution that has a
row $j$ as the most-right-ending row, for some $j$ that agrees
with $i$, is not equal to $k$ and ends before $i$. Adding row $i$
to the haplotype leads to an increase of its length of
$f(i,j)=r(i)-\max \{r(j),l(i)-1\}$. This term is fixed, for fixed
$i$ and $j$ and therefore we only have to consider extensions of
solutions that
were already optimal. Note that this reasoning does not hold for more general, ``gapped'', data.\\
\\
The last case is when $k=i$; $D[h,i;i]$ is equal to:
\begin{equation}
\max_{\substack{ j\in W(i),\text{ \ ~}j\neq h  \\ r(j)\leq r(i)}}%
\left\{
\begin{array}{l}
D[j,h;i-1]+f(i,j)\text{ if }r(h)\geq r(j),\\
D[h,j;i-1]+f(i,j)\text{ if }r(h)<r(j).%
\end{array}%
\right. \nonumber
\end{equation}
The above algorithm can be sped up by using the fact that, as a
direct consequence of (\ref{eq:lhr1}), $D[h,k;i]=D[h,k;max(h,k)]$
for all $h,k\leq i \leq n$. It is thus unnecessary to calculate
the
values $D[h,k;i]$ for $h,k<i$.\\
\\
The time for calculating all the $W(i)$ is $O(n^{2}m)$. When all
the $W(i)$ are known, it takes $O(n^{3})$ time to calculate all
the $D[h,k;max(h,k)]$. This is because we need to calculate
$O(n^{2})$ values $D[i,k;i]$ and also $O(n^{2})$ values $D[h,i;i]$
that take $O(n)$ time each. This leads to an overall time
complexity of $O(n^{2}m+n^{3})$.\\
\end{proof}
\vspace{-12pt}
\subsection{1-gap LHR is NP-hard and APX-hard}
\label{subsec:lhrhard}
\noindent\textbf{Problem:} \emph{1-gap LHR}\\
\textbf{Input: } SNP matrix $M$ with at most one gap per row\\
\textbf{Output: } The value LHR(M), as defined earlier\\
\\
In this section we prove that 1-gap LHR is APX-hard (and thus also
NP-hard.) We prove this by demonstrating (indirectly) an
L-reduction from the problem CUBIC-MAX-INDEPENDENT-SET - the
problem of computing the maximum cardinality of an independent set
in a cubic graph - which is itself proven
APX-hard in \cite{alimontikann}.\\
\\
We do this in several steps. We first show an L-reduction from
\emph{Single Haplotype} LHR (SH-LHR), the version of LHR where
only one haplotype is used\footnote{More formally:- rows of the
input matrix $M$ must be removed until the remaining rows are
mutually non-conflicting. The length of the resulting single
haplotype, which we seek to maximise, is the number of columns
(amongst the remaining rows) that have at least one non-hole
entry.}, to LHR, such that the number of gaps per rows is
unchanged. We then show an L-reduction from
CUBIC-MAX-INDEPENDENT-SET to 2-gap SH-LHR. Then, using an
observation pertaining to the structure of cubic graphs, we show
how this reduction can be adapted to give an L-reduction from
CUBIC-MAX-INDEPENDENT-SET to 1-gap SH-LHR. This proves
the APX-hardness of 1-gap SH-LHR and thus (by transitivity of L-reductions) also 1-gap LHR.\\
\begin{lemma}
\label{lem:shequiv} SH-LHR is L-reducible to LHR, such that the
number of gaps per row is unchanged.\\
\end{lemma}
\begin{proof}
Let $M$ be the $n \times m$ input to SH-LHR. We may assume that
$M$ contains no duplicate rows, because duplicate rows are
entirely redundant when working with only one haplotype. We map
the SH-LHR input, $M$, to the $2n \times m$ LHR input, $M'$, by
taking each row of $M$ and making a copy of it. Informally, the
idea is that the influence of the second haplotype can be neutralised
by doubling the rows of the input matrix. Note that this construction
clearly preserves the maximum number of gaps per row.\\
\\
Now, let $SOL(M')$ be the set that contains all pairs of
haplotypes $(H_1,H_2)$ that can be induced by removing some rows
of $M'$, partitioning the remaining rows of $M'$ into two mutually
non-conflicting sets, and then reading off the two induced
haplotypes. Similarly, let $SOL(M)$ be the set that contains all
haplotypes $H$ that can be induced by removing some rows of $M$
(such that the remaining rows are mutually non-conflicting) and
then reading off the single, induced haplotype. Note the following
pair of observations, which both follow directly from the
construction of $M'$:
\begin{equation}
\label{eq:shl1} (H_1,H_2) \in SOL(M') \Rightarrow H_1, H_2 \in
SOL(M),
\end{equation}
\begin{equation}
\label{eq:shl2} H \in SOL(M) \Rightarrow (H,H) \in SOL(M').
\end{equation}
To satisfy the L-reduction we need to show how elements from
$SOL(M')$ are mapped back to elements of $SOL(M)$ in polynomial
time. So, let $(H_1, H_2)$ be any pair from $SOL(M')$. If $|H_1|
\geq |H_2|$ map the pair $(H_1,H_2)$ to $H_1$, otherwise to $H_2$.
This completes the L-reduction, and we now prove its correctness.
Central to this is the proof of the following:
\begin{equation}
\label{eq:dubbel} \text{\emph{SH-LHR}}(M) = \frac{1}{2}
\text{\emph{LHR}}(M').
\end{equation}
The fact that SH-LHR(M) $\geq \frac{1}{2} \text{\emph{LHR}}(M')$
follows immediately from (\ref{eq:shl1}) and the mapping described
above. (This lets us fulfil condition \ref{eq:L1}) of the
L-reduction definition, taking $\alpha=2$.) The fact that
SH-LHR(M) $\leq \frac{1}{2} \text{\emph{LHR}}(M')$ follows
because, by (\ref{eq:shl2}), every element in $SOL(M)$ is
guaranteed to have a counterpart in $SOL(M')$ which has a total length twice as large.\\
\\
We can fulfil condition (\ref{eq:L2}) of the L-reduction by taking
$\beta=\frac{1}{2}$. To see this, let $(H_1, H_2)$ be any pair
from $SOL(M')$, and (wlog) assume that $|H_1| \geq |H_2|$. Let
$r=\text{\emph{LHR}}(M')$, the distance of $(H_1,H_2)$ from
optimal is then:
\begin{equation}
r - (|H_1|+|H_2|) \geq r - 2|H_1|.
\end{equation}
Let $l=\text{\emph{SH-LHR}}(M)$, then:
\begin{equation}
\begin{array}{ll}
l - |H_1|&{}={}\frac{r}{2} - |H_1|\\
&{}={}\frac{1}{2} \bigg ( r-2|H_1| \bigg )\\
&{}\leq{}\frac{1}{2} \bigg (r - (|H_1|+|H_2|) \bigg).
\end{array}
\end{equation}
Thus, taking $\beta = \frac{1}{2}$ satisfies condition
(\ref{eq:L2}) of the L-reduction.\\
\end{proof}
\begin{lemma}
\label{lem:2gapAPX} 2-gap SH-LHR is APX-hard\\
\end{lemma}
\begin{proof}
We reduce from CUBIC-MAX-INDEPENDENT-SET. Let $G = (V,E)$ be the
undirected, cubic input to CUBIC-MAX-INDEPENDENT-SET. We direct
the edges of $G$ in the manner described by Observation
\ref{obs:orient}, to give $\overrightarrow{G} = (V,
\overrightarrow{E})$. Thus, every vertex of $\overrightarrow{G}$
is now out-out-in or in-in-out. A vertex $w$ is a \emph{child} of
a vertex $v$ if there is an edge leaving $v$ in the direction of
$w$ i.e. $(v,w) \in \overrightarrow{E}$, and in this case
$v$ is said to be the \emph{parent} of $w$.\\
\\
Let $v_{in}$ be the number of vertices in $\overrightarrow{G}$
that are in-in-out, and $v_{out}$ be the number of vertices that
are out-out-in. We build a matrix $M$, to be used as input to
2-gap SH-LHR, which has $|V|$ rows and $2v_{in} + v_{out}$
columns. The construction of $M$ is as follows. (Each row of $M$
will represent a vertex from $V$, so we henceforth index the rows
of $M$ using vertices of $V$.) Now, to each in-in-out vertex of
$\overrightarrow{G}$, we allocate two \emph{adjacent} columns of
$M$, and for each out-out-in vertex, we allocate one column of
$M$. (A column may not be allocated to more than one
vertex.)\footnote{Note that, for this lemma, it is not important
how the columns are allocated; in the proof of Lemma
\ref{lem:lhrhard}, the ordering is crucial.} For simplicity, we
also impose an arbitrary total order
$P$ on the vertices of $V$.\\
\\
Now, for each vertex $v \in V$, we build row $v$ as follows.
Firstly, we put 1(s) in the column(s) representing $v$. Secondly,
consider each child $w$ of $v$. If $w$ is an out-out-in vertex, we
put a $0$ in the column representing $w$. Alternatively, $w$ is an
in-in-out vertex, so $w$ is represented by two columns; in this
case we put a 0 in the left such column (if $v$ comes before the
other parent of $w$ in the total order $P$) or, alternatively, in
the right column (if $v$ comes after the other parent of $w$ in the
total order $P$). The rest of the row is holes.\\
\begin{figure}
\begin{center}
\epsfig{file=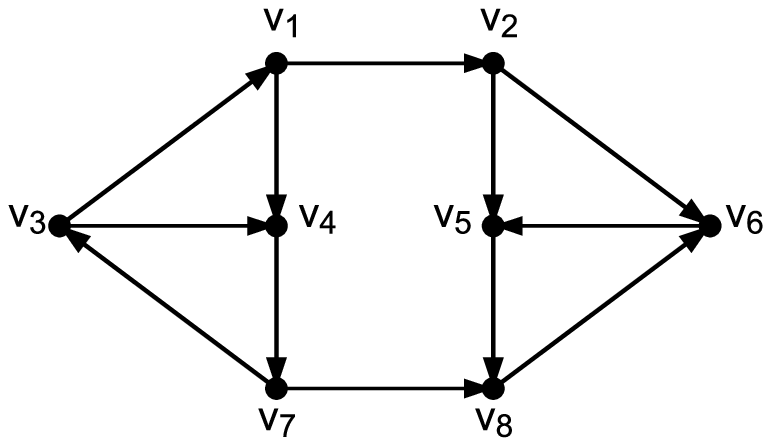}
\end{center}
\caption{Example input graph to CUBIC-MAX-INDEPENDENT-SET (see
Lemmas \ref{lem:2gapAPX} and \ref{lem:shlhrhard}) after an
appropriate edge orientation has been applied.}
\label{fig:lhrgraph}
\end{figure}
\newlength{\blb}
\setlength{\blb}{-3.0pt}
\newlength{\bl}
\setlength{\bl}{-4pt}
\newlength{\lb}
\setlength{\lb}{-12.0pt}
\begin{figure}
\begin{center}
\begin{tabular}{l}
$\left.
\begin{array}{ccccccccccccc}
\hspace{28pt} & v_3 \hspace{\bl} & v_1 \hspace{\bl} & v_2
\hspace{\bl} & v_5 \hspace{\bl} & v_5 \hspace{\bl} & v_7
\hspace{\bl} & v_8 \hspace{\bl} & v_8 \hspace{\bl} &
v_4 \hspace{\bl} & v_4 \hspace{\bl} & v_6 \hspace{\bl} & v_6%
\end{array}
\right.$\\
\begin{tabular}{ll}
$\left.
\begin{array}{c}
v_1\\
v_2\\
v_3\\
v_4\\
v_5\\
v_6\\
v_7\\
v_8%
\end{array}
\right. $
&
$\hspace{\lb}\left(
\begin{array}{cccccccccccc}
- \hspace{\blb} & 1 \hspace{\blb} & 0 \hspace{\blb} & - \hspace{\blb} & - \hspace{\blb} & - \hspace{\blb} & - \hspace{\blb} & - \hspace{\blb} & 0 \hspace{\blb} & - \hspace{\blb} & - \hspace{\blb} & - \\
- \hspace{\blb} & - \hspace{\blb} & 1 \hspace{\blb} & 0 \hspace{\blb} & - \hspace{\blb} & - \hspace{\blb} & - \hspace{\blb} & - \hspace{\blb} & - \hspace{\blb} & - \hspace{\blb} & 0 \hspace{\blb} & - \\
1 \hspace{\blb} & 0 \hspace{\blb} & - \hspace{\blb} & - \hspace{\blb} & - \hspace{\blb} & - \hspace{\blb} & - \hspace{\blb} & - \hspace{\blb} & - \hspace{\blb} & 0 \hspace{\blb} & - \hspace{\blb} & - \\
- \hspace{\blb} & - \hspace{\blb} & - \hspace{\blb} & - \hspace{\blb} & - \hspace{\blb} & 0 \hspace{\blb} & - \hspace{\blb} & - \hspace{\blb} & 1 \hspace{\blb} & 1 \hspace{\blb} & - \hspace{\blb} & - \\
- \hspace{\blb} & - \hspace{\blb} & - \hspace{\blb} & 1 \hspace{\blb} & 1 \hspace{\blb} & - \hspace{\blb} & - \hspace{\blb} & 0 \hspace{\blb} & - \hspace{\blb} & - \hspace{\blb} & - \hspace{\blb} & - \\
- \hspace{\blb} & - \hspace{\blb} & - \hspace{\blb} & - \hspace{\blb} & 0 \hspace{\blb} & - \hspace{\blb} & - \hspace{\blb} & - \hspace{\blb} & - \hspace{\blb} & - \hspace{\blb} & 1 \hspace{\blb} & 1 \\
0 \hspace{\blb} & - \hspace{\blb} & - \hspace{\blb} & - \hspace{\blb} & - \hspace{\blb} & 1 \hspace{\blb} & 0 \hspace{\blb} & - \hspace{\blb} & - \hspace{\blb} & - \hspace{\blb} & - \hspace{\blb} & - \\
- \hspace{\blb} & - \hspace{\blb} & - \hspace{\blb} & - \hspace{\blb} & - \hspace{\blb} & - \hspace{\blb} & 1 \hspace{\blb} & 1 \hspace{\blb} & - \hspace{\blb} & - \hspace{\blb} & - \hspace{\blb} & 0 %
\end{array}
\right)$
\end{tabular}
\end{tabular}
\caption{Construction of matrix $M$ (from Lemma \ref{lem:2gapAPX}
and \ref{lem:shlhrhard}) for graph in Figure \ref{fig:lhrgraph}}
\label{fig:lhrmatrix}
\end{center}
\vspace{-12pt}
\end{figure}
\\
This completes the construction of $M$. Note that rows encoding
in-in-out vertices contain two adjacent 1s and one 0, with at most
one gap in the row, and rows encoding out-out-in vertices contain
one 1 and two 0s, with at most two gaps in the row. In either case
there are precisely 3 non-hole elements per row. It is also
crucial to note that, reading
down any one column of $M$, one sees exactly one 1 and exactly one 0.\\
\\
Let $K$ be any submatrix of $M$ obtained by removing rows from
$M$, and let $V[K] \subseteq V$ be the set of vertices whose rows
appear in $K$. If the rows of $K$ are mutually non-conflicting,
then the haplotype induced by $K$ has length $3r$ where $r$ is the
number of rows in $K$. This follows from the aforementioned facts
that every column of $M$ contains exactly one 1 and
one 0. and that every row has exactly 3 non-hole elements.\\
\\
We now prove that the rows of $K$ are in conflict iff $V[K]$ is
not an independent set. First, suppose $V[K]$ is not an
independent set. Then there exist $u, v \in V[K]$ such that $(u,v)
\in \overrightarrow{E}$. In row $v$ of $K$ there are thus 1(s) in
the column(s) representing vertex $v$. However, there is also (in
row $u$) a 0 in the column (or one of the columns) representing
vertex $v$, causing a conflict. Hence, if $V[K]$ is not an
independent set, $K$ is in conflict. Now consider the other
direction. Suppose $K$ is in conflict. Then in some column of $K$
there is a 0 and a 1. Let $u$ be the row where the 0 is seen, and
$v$ be the row where the 1 is seen. So both $u$ and $v$ are in
$V[K]$. Further, we know that there is an out-edge $(u,v)$ in
$\overrightarrow{E}$, and thus an edge between $u$ and $v$ in $E$,
proving
that $V[K]$ is not an independent set. This completes the proof of the iff relationship.\\
\\
It follows that:
\begin{equation}
\begin{array}{l}
\text{\emph{CUBIC-MAX-INDEPENDENT-SET}}(G)\\
= \frac{1}{3} \text{\emph{SH-LHR}}(M).
\end{array}
\end{equation}
The conditions of the L-reduction definition are now easily
satisfied, because of the 1-1 correspondence between haplotypes
induced (after row-removals) and independent sets in $G$, and the
fact that a size-$r$ independent set of $G$ corresponds to a
length-$3r$ haplotype (or, equivalently, to $r$ mutually
non-conflicting rows of $M$.) The L-reduction is formally
satisfied by taking $\alpha = 3$ and $\beta = \frac{1}{3}$. The
two functions that comprise the L-reduction are both polynomial
time computable.\\
\end{proof}
\begin{lemma}
\label{lem:shlhrhard} 1-gap SH-LHR is APX-hard.\\
\end{lemma}
\begin{proof}
This proof is almost identical to the proof of Lemma
\ref{lem:2gapAPX}; the difference is the manner in which columns
of $M$ are assigned to vertices of $G$. The informal motivation is
follows. In the previous allocation of columns to vertices, it was
possible for a row corresponding to an out-out-in vertex to have 2
gaps. Suppose, for each out-out-in vertex, we could ensure that
one of the 0s in its row was adjacent to the 1 in the row, with no
holes in between. Then every row of the matrix would have (at
most) 1 gap, and we would be finished. We now show that, by
exploiting a rather subtle property
of cubic graphs, it is indeed possible to allocate columns to vertices such that this is possible.\\
\\
Assume, that we have ordered the edges of $G$ as before to obtain
$\overrightarrow{G}$. Let $V_{out} \subseteq V$ be those vertices
in $V$ that are out-out-in. Now, suppose we could compute (in
polynomial time) an injective function $favourite: V_{out}
\rightarrow V$ with the following properties:
\begin{itemize}
\item for every $v \in V_{out}$, $(v,favourite(v))\in
\overrightarrow{E}$;
% \item for every $u,v \in V_{out}$, $favourite(u) = favourite(v)$
% iff $u=v$; % follows from the fact that the function is injective
\item the subgraph of $\overrightarrow{G}$ induced by edges of the
form $(v,favourite(v))$, henceforth called the
\emph{favourite-induced subgraph}, is acyclic.\\
\end{itemize}
Given such a function it is easy to create a total enumeration of
the vertices of $V$ such that every out-out-in vertex is
immediately followed by its \emph{favourite} vertex. This
enumeration can then be used to allocate the columns of $M$ to the
vertices of $V$, such that every row of $M$ has at most one gap.
To ensure this property, it is necessary to stipulate that, where
$favourite(v)$ is an in-in-out vertex, the 0 encoding the edge
$(v,favourite(v))$ is placed in the \emph{left} of the two columns
encoding $favourite(v)$. This is not a problem because every
vertex is the favourite of at most one other vertex.\\
\\
It remains to prove that the function \emph{favourite} exists and
that it can be constructed in polynomial time. This is equivalent
to finding vertex disjoint directed paths in $\overrightarrow{G}$
such that every out-out-in vertex is on such a path and all paths
end in an in-in-out vertex. Lemma \ref{lem:bert} tells us how to
find such paths. We thank Bert Gerards for invaluable
help with this.\\
\\
This completes the proof that 1-gap SH-LHR is APX-hard. (See
Figures \ref{fig:lhrgraph} and \ref{fig:lhrmatrix} for
an example of the whole reduction in action.)\\
\end{proof}
\begin{lemma}
\label{lem:bert} Let $\overrightarrow{G}$ be a directed, cubic
graph with a partition $(V_{out},V_{in})$ of the vertices such
that the vertices in $V_{out}$ are out-out-in and the vertices in
$V_{in}$ are in-in-out. Then $V_{out}$ can be covered, in
polynomial time, by vertex-disjoint directed paths ending in
$V_{in}$.\\
\end{lemma}
\begin{proof}
Observe that any two directed circuits contained entirely within
$V_{out}$ are pairwise vertex disjoint. Let $V'_{out}$ be obtained
from $V_{out}$ by shrinking each directed circuit in $V_{out}$ to
a single vertex, and let $\overrightarrow{G'}$ be the resulting
new graph. (Note that each vertex in $V'_{out}$ has outdegree at
least 2 and indegree at most 1 and that the indegree of each node
in $V_{in}$ is still 2, because we do not delete multiple edges)
We now argue that it is possible to find a set of edges $F'$ in
$\overrightarrow{G'}$, with $|F'| = |V'_{out}|$, such that - for
each $v \in V'_{out}$ - precisely one edge from $F'$ begins at
$v$, and such that no two edges in $F'$ have the same endpoint. We
prove this by construction. For each vertex $u \in V'_{out}$ that
has a child $v$ in $V'_{out}$, we can add the edge $(u,v)$ to
$F'$, because $v$ has indegree 1 and therefore no other edges can
end at $v$. (In case $u$ has two such children, we can choose one
of the edges to add to $F'$). Thus we are left to deal with a
subset of vertices $L \subseteq V'_{out}$ where every vertex in
$L$ has all its children in $V_{in}$. Now consider the bipartite
graph $B$ with bipartition $(L, V_{in})$ and an edge for every
directed edge of $\overrightarrow{G'}$ going from $L$ to $V_{in}$.
If we can find a matching in $B$ of size $|L|$, we can complete
the construction of $F'$ by adding the edges from the perfect
matching. Hall's Theorem states that a bipartite graph with
bipartition $(X,Y)$ has a matching of size $|X|$ iff, for all $X'
\subseteq X$, $|N(X')| \geq |X'|$, where $N(X')$ is the set of all
neighbours of $X'$. Now, note that each vertex in $L$ sends at
least two edges across the partition of $B$, and each vertex in
$V_{in}$ can accept at most two such edges, so for each $L'
\subseteq L$ it is clear that $|N(L')| \geq |L'|$. Hence, the
graph $(L, V_{in})$ does indeed have a matching of size $|L|$ and
the construction of $F'$ can be completed.\\
\\
Now, given that the graph induced by $V'_{out}$ is acyclic, so is
$F'$. Let $F$ be the set of edges in $\overrightarrow{G}$
corresponding to those in $F'$. $F$ is acyclic and each directed
circuit $C$ in $V_{out}$ has exactly one vertex $v_{C}$ that is a
tail of an edge of $F$ and no vertex that is a head of an edge in
$F$. Let $P_C$ be the longest directed path in $C$ that ends in
$v_C$. Then the union of $F$ and all $P_C$ over all directed
circuits $C$ in $V_{out}$ is a collection of paths ending in
$V_{in}$ and covering $V_{out}$.\\
\\
Finding cycles in a graph and finding a maximum matching in a
bipartite graph are both polynomial-time computable, so the whole
process described above is polynomial-time computable.\\
\end{proof}
\begin{lemma}
\label{lem:lhrhard}
1-gap LHR is APX-hard.\\
\end{lemma}
\begin{proof}
Follows from Lemma \ref{lem:shlhrhard} and Lemma
\ref{lem:shequiv}.\\
\end{proof}
\section{Conclusion}
This paper involves the complexity (under various different input
restrictions) of the haplotyping problems Minimum Error Correction
(MEC) and Longest Haplotype Reconstruction (LHR). The state of
knowledge about MEC and LHR after this paper is demonstrated in
Table \ref{tab:after}. We also include Minimum Fragment Removal
(MFR) and Minimum SNP Removal (MSR) in the table because they are
two other well-known Single Individual Haplotyping problems. MSR
(MFR) is the problem of removing the minimum number of columns
(rows) from an SNP-matrix in order to make it feasible.\\
\begin{table}[h]
\begin{centering}
\begin{tabular}{|c||c|c|}
\hline
& Binary (i.e. no holes) & ? (Section \ref{subsec:bmec})\\
& & PTAS known \cite{li}\\
\cline{2-3}
MEC & Ungapped & NP-hard (Section \ref{subsec:umec})\\
\cline{2-3}
 & 1-Gap & NP-hard (Section \ref{subsec:gmec}),\\
 & & APX-hard (Section \ref{subsec:gmec})\\
% \cline{2-3}
%  & General & NP-hard (implicit in \cite{kleinberg})\\
%  & & APX-hard (Section \ref{subsec:gmec})\\
\hline
%
% & Binary (i.e. no holes) & P (trivially)\\
% \cline{2-3}
 & Ungapped & P (Section \ref{subsec:lhrpoly})\\
\cline{2-3}
LHR & 1-Gap & NP-hard (Section \ref{subsec:lhrhard})\\
 & & APX-hard (Section \ref{subsec:lhrhard})\\
% \cline{2-3}
%  & General & NP-hard (Section \ref{subsec:lhrhard})\\
%  & & APX-hard (Section \ref{subsec:lhrhard})\\
\hline
& Ungapped & P \cite{bafna2005}\\
\cline{2-3}
MFR & 1-Gap & NP-hard \cite{lanciabafna}\\
 & & APX-hard \cite{bafna2005}\\
% \cline{2-3}
%  & General & NP-hard \cite{lanciabafna}\\
%  & & APX-hard \cite{bafna2005}\\
\hline
& Ungapped & P \cite{lanciabafna}\\
\cline{2-3}
MSR & 1-Gap & NP-hard \cite{bafna2005}\\
 & & APX-hard \cite{bafna2005}\\
% \cline{2-3}
%  & General & NP-hard \cite{lanciabafna}\\
%  & & APX-hard \cite{bafna2005}\\
\hline
\end{tabular}
\caption{The new state of knowledge following our work}
\label{tab:after}
\end{centering}
\vspace{-12pt}
\end{table}
\\
Indeed, from a complexity perspective, the most intriguing open
problem is to ascertain the complexity of the ``re-opened''
problem Binary-MEC. It would also be interesting to study the
approximability of Ungapped-MEC.\\
% ; we conjecture that (in an
% approximation complexity sense) it is somewhat easier
% than 1-gap MEC.\\
\\
From a more practical perspective, the next logical step is to
study the complexity of these problems under more restricted
classes of input, ideally under classes of input that have direct
biological relevance. It would also be of interest to study some
of these problems in a ``weighted'' context i.e. where the cost of
the operation in question (row removal, column removal, error
correction) is some function of (for example) an \emph{a priori}
specified confidence in the correctness of the data being changed.
\section{Acknowledgements}
We thank Leen Stougie and Judith Keijsper for many useful
conversations during the writing of this paper.
%
%
%
       % end the bibliography
%
%
\vspace{30pt}
\begin{biography}{Rudi Cilibrasi}
received his bachelor's degree at Caltech in 1996. He spent
several years in industry doing network programming, Linux kernel
programming, and a variety of software development work until
returning to academia with CWI in 2001. He is now nearing
completion of his doctoral work that has been largely concerned
with robust methods of approximating bioinformatics and related
clustering problems. He currently maintains CompLearn (
http://complearn.org/ ), an open-source data-mining package that
can be used for phylogenetic tree construction.

\end{biography}
\begin{biography}{Leo van Iersel}
received in 2004 his Master of Science degree in Applied
Mathematics from the Universiteit Twente in the Netherlands. He is
now working as a PhD student at the Technische Universiteit
Eindhoven, also in the Netherlands. His research is mainly
concerned with the search for combinatorial algorithms for
biological problems.
\end{biography}
\begin{biography}{Steven Kelk}
received his PhD in Computer Science in 2004 from the University
of Warwick, in England. He is now working as a postdoc at the
Centrum voor Wiskunde en Informatica (CWI) in Amsterdam, the
Netherlands, where he is focussing on the combinatorial aspects of
computational biology.
\end{biography}
\begin{biography}{John Tromp}
received the bachelor's and PhD degrees in Computer Science from
the University of Amsterdam in 1989 and 1993 respectively, where
he studied with Paul Vit\'{a}nyi. He then spent two years as a
postdoctoral fellow with Ming Li at the University of Waterloo in
Canada. In 1996 he returned as a postdoc to the Centre for
Mathematics and Computer Science (CWI) in Amsterdam. He spent 2001
working as software developer at Bioinformatics Solutions Inc. in
Waterloo, to return once more to CWI, where he currently holds a
permanent position. He is the recipient of a Canada International
Fellowship. See http://www.cwi.nl/~tromp/ for more information.
\end{biography}
\clearpage
\section*{Appendix: Interreducibility of MEC and Constructive-MEC}
\label{app:inter}
\renewcommand{\theequation}{A\arabic{equation}}
\setcounter{section}{0}
\renewcommand{\thesection}{A.\arabic{section}}
\numberwithin{equation}{section} \numberwithin{figure}{section}
%
% \section{Interreducibility of MEC and Constructive-MEC}
%
\begin{lemma}
\label{lem:int} MEC and Constructive-MEC are polynomial-time
Turing interreducible. (Also: Binary-MEC and
Binary-Constructive-MEC are polynomial-time Turing
interreducible.)\\
\end{lemma}
\begin{proof}
We show interreducibility of MEC and Constructive-MEC in such a
way that the interreducibility of Binary-MEC with
Binary-Constructive-MEC also follows immediately from the
reduction. This makes the reduction from Constructive-MEC to
MEC quite complicated because we must thus avoid the use of holes.\\
\\
1. Reducing MEC to Constructive-MEC is trivial because, given an
optimal haplotype pair $(H_1, H_2)$, $D_M(H_1, H_2)$ can easily be
computed in polynomial-time by summing $\min( d(H_1,r), d(H_2,r)
)$ over all rows $r$ of the
input matrix $M$.\\
\\
2. Reducing Constructive-MEC to MEC is more involved. To prevent a
particular special case which could complicate our reduction, we
first check whether every row of $M$ (i.e. the input to
Constructive-MEC) is identical. If this is so, we can complete the
reduction by simply returning $(H_1, H_1)$ where $H_1$ is the
first row of $M$. Hence,
from this point onwards, we assume that $M$ has at least two distinct rows.\\
\\
Let $OptPairs(M)$ be the set of all unordered optimal haplotype
pairs for $M$ i.e. the set of all $(H_1, H_2)$ such that $D_M(H_1,
H_2) = MEC(M)$. Given that all rows in $M$ are not identical, we
observe that there are no pairs of the form $(H_1, H_1)$ in
$OptPairs(M)$.\footnote{This is because $D_{M}(H_1,H_1)$ is always
larger than $D_{M}(H_1, r)$ for any row $r$ in $M$ that is not
equal to $H_1$.} Let $OptPairs(M,H') \subseteq OptPairs(M)$ be
those elements $(H_1,H_2) \in OptPairs(M)$ such that $H_1 = H'$ or
$H_2 = H'$. Let $g(r, H_1, H_2)$
be defined as $\min( d(r, H_1), d(r, H_2) )$.\\
\\
Consider the following two subroutines:\\
\\
\textbf{Subroutine: } \emph{DFN} (``Distance From Nearest Optimal Haplotype Pair'')\\
\textbf{Input: } An $n \times m$ SNP matrix $M$ and a vector $r \in \{0,1\}^m$.\\
\textbf{Output: } The value $d_{dfn}$ which we define as follows:
\begin{equation}
d_{dfn} = \min_{ (H_1, H_2) \in OptPairs(M) } g( r, H_1, H_2
).\nonumber
\end{equation}
\\
\textbf{Subroutine: } \emph{ANCHORED-DFN} (``Anchored Distance From Nearest Optimal Haplotype Pair'')\\
\textbf{Input: } An $n \times m$ SNP matrix $M$, a vector $r \in
\{0,1\}^m$, and a haplotype $H'$ such that
$(H', H_2) \in OptPairs(M)$ for some $H_2$.\\
\textbf{Output: } The value $d_{adfn}$, defined as:
\begin{equation}
d_{adfn} = \min_{ (H_1, H_2) \in OptPairs(M, H') } g( r, H_1, H_2
).\nonumber
\end{equation}
\\
We assume the existence of implementations of DFN and ANCHORED-DFN
which run in polynomial-time whenever MEC runs in polynomial-time.
We use these two subroutines to reduce Constructive-MEC to MEC and
then, to complete the proof, demonstrate and prove correcteness of
implementations for
DFN and ANCHORED-DFN.\\
\\
The general idea of the reduction from Constructive-MEC to MEC is
to find some pair $(H_1, H_2) \in OptPairs(M)$ by first finding
$H_1$ (using repeated calls to DFN) and then finding $H_2$ (by
using repeated calls to ANCHORED-DFN with $H_1$ specified as the
``anchoring'' haplotype.) Throughout the reduction, the following
two observations are important. Both follow immediately from the
definition of $D$ - i.e. (\ref{eq:witsum}).\\
\begin{observation}
\label{obs:expand} Let $M_1 \cup M_2$ be a partition of rows of
the matrix $M$ into two sets. Then, for all $H_1$ and $H_2$,
$D_{M}(H_1, H_2) = D_{M_1}(H_1, H_2) + D_{M_2}(H_1,H_2)$.\\
\end{observation}
\begin{observation}
\label{obs:baseline} Suppose an SNP matrix $M_1$ can be obtained
from an SNP matrix $M_2$ by removing 0 or more rows from $M_2$.
Then $MEC(M_1) \leq MEC(M_2)$.\\
\end{observation}
To begin the reduction, note that, for an arbitrary
haplotype $X$, DFN$(M,X)=0$ iff $(X, H_2) \in OptPairs(M)$ for
some haplotype $H_2$. Our idea is thus that we initialise $X$ to
be all-0 and flip one entry of $X$ at a time (i.e. change a 0 to a
1 or vice-versa) until DFN$(M,X)=0$; at that point $X = H_1$ (for
some $(H_1, H_2) \in OptPairs(M)$.) More specifically, suppose
DFN$(M,X) = d$ where $0 < d < m$. \footnote{It is not possible
that DFN$(M,X)=m$, because all $(H_1, H_2) \in OptPairs(M)$ are of
the form $H_1 \neq H_2$, and if $H_1 \neq H_2$ we know that
$g(X,H_1,H_2) < m$.} If we define $flip(X,i)$ as the haplotype
obtained by flipping the entry in the $i$th column of $X$, then we
know that there exists $i$ ($1 \leq i \leq m$) such that DFN$(M,
flip(X,i)) < d$. Such a position must exist because we can flip
some entry in $X$ to bring it closer to the haplotype (which we
know exists) that it was distance $d$ from. It is clear that we
can find a position $i$ in polynomial-time by calling DFN$(M,
flip(X,j))$ for $1 \leq j \leq m$ until it is found.
Having found such an $i$, we set $X = flip(X,i)$.\\
\\
Clearly this process can be iterated, finding one entry to flip in
every iteration, until DFN$(M,X)=0$ and at this point setting $H_1
= X$ gives us the desired result. Given that DFN$(M,X)$ decreases
by at least 1 every iteration, at most $m-1$ iterations
are required.\\
\\
Thus, having found $H_1$, we need to find some $H_2$ such that
$(H_1, H_2)$ is in $OptPairs(M)$.\\
\\
First, we initialise $X$ to be the complement of $H_1$ (i.e. the
row obtained by flipping every entry of $H_1$). Now, observe that
if $X \neq H_1$ and ANCHORED-DFN$(M, X, H_1) = 0$ then $(H_1, X)
\in OptPairs(M)$ and we are finished. The tactic is thus to find,
at each iteration, some position $i$ of $X$ such that
ANCHORED-DFN$(M, flip(X,i), H_1)$ is less than ANCHORED-DFN$(M,
X,H_1)$, and then setting $X$ to be $flip(X,i)$. As before we
repeat this process until our call to ANCHORED-DFN returns zero.
The ``trick'' in this case is to prevent $X$ converging on $H_1$,
because (knowing that $M$ has at least two different types of row)
$(H_1, H_1) \not \in OptPairs(M)$. The initialisation of $X$ to
the complement of $H_1$ guarantees this. To see why this is,
observe that, if $X$ is the complement of $H_1$, $d(X,H_1)=m$.
Thus, we would need at least $m$ flips to transform $X$ into
$H_1$. However, if $X$ is the complement of $H_1$, then - because
we have guaranteed that $OptPairs(M)$ contains no pairs of the
form $(H_1,H_1)$ - we know that ANCHORED-DFN$(M, X, H_1) < m$.
Given that we can guarantee that ANCHORED-DFN$(M,X,H_1)$ can be
reduced by at least 1 at every iteration, it is clear that we can
find an $X$ such that ANCHORED-DFN$(M,X,H_1)=0$ after making no
more than $m-1$ iterations, which ensures that $X$ cannot have
been transformed into $H_1$. Once we have such an $X$ we can set
$H_2 = X$ and
return $(H_1, H_2)$.\\
\\
To complete the proof of Lemma \ref{lem:int} it remains only to
demonstrate and prove the correctness of algorithms for DFN and
ANCHORED-DFN, which we do below. Note that both DFN and
ANCHORED-DFN run in polynomial-time if MEC runs in
polynomial-time.\\
\\
\textbf{Subroutine: } \emph{DFN} (``Distance From Nearest Optimal Haplotype Pair'')\\
\textbf{Input: } An $n \times m$ SNP matrix $M$ and a vector $r \in \{0,1\}^m$.\\
\textbf{Output: } The value $d_{dfn}$ which we define as follows:
\begin{equation}
d_{dfn} = \min_{ (H_1, H_2) \in OptPairs(M) } g( r, H_1, H_2
).\nonumber
\end{equation}
The following is a three-step algorithm to compute DFN(M,r) which uses an oracle for MEC.\\
\\
1. Compute $d = $MEC$(M)$.\\
2. Let $M'$ be the $n(m+1) \times m$ matrix obtained from $M$ by
making $m+1$ copies of every row of $M$.\\
3. Return MEC$( M' \cup \{r\} ) - (m+1)d$ where $M' \cup \{r\}$ is
the matrix obtained by adding the single row $r$ to the matrix
$M'$.\\
\\
To prove the correctness of the above we first make a further
observation, which (as with the two previous observations) follows
directly from (\ref{eq:witsum}).\\
\begin{observation}
\label{obs:scale} Suppose an $kn \times m$ SNP matrix $M_1$ is
obtained from an $n \times m$ SNP matrix $M_2$ by making $k \geq
1$ copies of every row of $M_2$. Then $MEC(M_1) = k.MEC(M_2)$, and
$OptPairs(M_1) = OptPairs(M_2)$.\\
\end{observation}
By the above observation we know that MEC$(M') = (m+1)d$ and
$OptPairs(M') = OptPairs(M)$. Now, we argue that $OptPairs(M' \cup
\{r\}) \subseteq OptPairs(M)$. To see why this is, suppose there
existed $(H_3, H_4)$ such that $(H_3, H_4) \in OptPairs(M' \cup
\{r\})$ but $(H_3, H_4) \not \in OptPairs(M)$. This would mean
$D_{M}(H_3, H_4) > d$ where $d = $MEC$(M)$. Now:
\begin{align*}
D_{M' \cup \{r\}}(H_3, H_4) & \geq D_{M'}(H_3, H_4)\\
& = (m+1)D_{M}(H_3, H_4)\\
& \geq (m+1)(d+1).
\end{align*}
However, if we take any $(H_1, H_2) \in OptPairs(M)$, we see that:
\begin{align*}
D_{M' \cup \{r\}}(H_1, H_2) & \leq (m+1)d + g(r,H_1, H_2)\\
& \leq (m+1)d + m.
\end{align*}
Now, $(m+1)d + m < (m+1)(d+1)$ so $(H_3, H_4)$ could not possibly
be in $OptPairs(M' \cup \{r\})$ - contradiction! The relationship
$OptPairs(M' \cup \{r\}) \subseteq OptPairs(M)$ thus follows. It
further follows, from Observation \ref{obs:expand}, that the
members of $OptPairs(M' \cup \{r\})$ are precisely those pairs
$(H_1, H_2) \in OptPairs(M)$ that minimise the expression
$g(r,H_1,H_2)$. The minimal value of $g(r, H_1, H_2)$ has already
been defined as $d_{dfn}$, so we have:
\begin{equation}
MEC(M' \cup \{r\}) = (m+1)d + d_{dfn}.\nonumber
\end{equation}
This proves the correctness of Step 3 of the subroutine.\\
\\
\textbf{Subroutine: } \emph{ANCHORED-DFN} (``Anchored Distance From Nearest Optimal Haplotype Pair'')\\
\textbf{Input: } An $n \times m$ SNP matrix $M$, a vector $r \in
\{0,1\}^m$, and a haplotype $H'$ such that
$(H', H_2) \in OptPairs(M)$ for some $H_2$.\\
\textbf{Output: } The value $d_{adfn}$, defined as:
\begin{equation}
d_{adfn} = \min_{ (H_1, H_2) \in OptPairs(M, H') } g( r, H_1, H_2
).\nonumber
\end{equation}
Given that $H'$ is one half of some optimal haplotype pair for
$M$, it can be shown that ANCHORED-DFN$(M, r, H')$ =  DFN$( M \cup
\{H'\}, r)$, thus demonstrating how ANCHORED-DFN can be easily
reduced to DFN in polynomial-time. To prove the equation it is
sufficient to demonstrate that $OptPairs( M \cup \{H'\}) =
OptPairs(M, H')$, which we do now. Let $d=$MEC$(M)$. It follows
that MEC$( M \cup \{ H' \} ) \geq d$. In fact, MEC$(M \cup \{H'\})
= d$ because $D_{M \cup \{H'\}}( H', H_2 ) = d$ for all $(H', H_2)
\in OptPairs(M,H')$. Hence $OptPairs(M,H') \subseteq OptPairs(M
\cup \{ H' \})$. To prove the other direction, suppose there
existed some pair $(H_1, H_2) \in OptPairs(M \cup \{ H' \})$ such
that $H_1 \neq H'$ and $H_2 \neq H'$. But then, from Observation
\ref{obs:expand}, we would have:
\begin{align*}
D_{M \cup \{H' \}} (H_1, H_2) &= D_{M}(H_1, H_2) + g(H', H_1, H_2) \\
&\geq D_{M}(H_1, H_2) + 1\\
&> d.
\end{align*}
Thus, $(H_1, H_2)$ could not have been in $OptPairs(M \cup
\{H'\})$ in the first place, giving us a contradiction. Thus
$OptPairs(M \cup \{ H' \}) \subseteq OptPairs(M, H')$ and hence
$OptPairs(M \cup \{H' \}) = OptPairs(M, H')$, proving the
correctness of subroutine ANCHORED-DFN.
\end{proof}
\end{document}